\documentclass[a4paper,10.5pt]{article}

\usepackage[dvips]{graphicx}
\usepackage{float}
\usepackage{epsfig,subfigure}
\usepackage{longtable}
\usepackage{rotating}
\usepackage{amsmath}
\usepackage{amsfonts}
\usepackage{amssymb}
\usepackage{cite}
\usepackage{dcolumn}
\usepackage{bm}
\usepackage{hyperref}   

\usepackage{color}
\usepackage{txfonts}

\newcommand{\sqrts}{\sqrt{s}}
\newcommand{\sqrtsnn}{\sqrt{s_{_{\mbox{\rm \tiny{NN}}}}}}
\newcommand{\alphaS}{\alpha_{\rm s}}
\newcommand\cO{{\cal O}}
\newcommand\cN{{\cal N}}
\newcommand\bR{{\cal B}}

\newcommand{\Sig}{\mathcal{S}}

\newcommand{\LumiInt}{\mathcal{L}_{\mbox{\rm \tiny{int}}}}

\newcommand{\pp}{{\rm{p-p}}}
\newcommand{\pn}{{\rm{p-n}}}
\newcommand{\ppbar}{{\rm{p-$\bar{\rm p}$}}}
\newcommand{\pA}{{\rm{p-A}}}
\newcommand{\pN}{{\rm{p-N}}}
\newcommand{\pPb}{{\rm{p-Pb}}}
\newcommand{\PbPb}{{\rm{Pb-Pb}}}
\newcommand{\TpA}{T_{_{\rm pA}}}

\newcommand{\TpAsq}{T^{2}_{_{\rm pA}}}

\newcommand{\dtwor}{{\rm{d^2r}}}

\newcommand{\pythia}{{\sc pythia}}
\newcommand{\herwig}{{\sc herwig}}
\newcommand{\mcfm}{{\sc mcfm}}
\newcommand{\vbfnlo}{{\sc vbfnlo}}

\newcommand{\sigmaDPS}{\sigma^{{\rm {\tiny DPS}}}}

\newcommand{\sigmaDPSone}{\sigma^{{\rm {\tiny DPS,1}}}}
\newcommand{\sigmaDPStwo}{\sigma^{{\rm {\tiny DPS,2}}}}
\newcommand{\sigmaSPS}{\sigma^{{\rm {\tiny SPS}}}}
\newcommand{\sigmaeff}{\sigma_{\rm eff}}
\newcommand{\sigmaeffpp}{\sigma_{_{\rm eff,pp}}}
\newcommand{\sigmaeffpA}{\sigma_{_{\rm eff,pA}}}

\newcommand{\ttbar}    {\ensuremath{t\bar{t}}}
\newcommand{\MET}{\ensuremath{{E\!\!\!/}_{\mathrm{T}}}}

\def\ttt#1{\texttt{\scriptsize #1}}

\begin{document}

\begin{center}
\vbox to 1 truecm {}

{\Large \bf Same-sign WW production in proton-nucleus collisions}
\\[0.2cm] 
{\Large \bf at the LHC as a signal for double parton scattering} \\[0.8cm]

{\large David~d'Enterria$^{a}$ and Alexander~M.~Snigirev$^b$}\\[0.5cm]

{\it $^a$ CERN, PH Department, 1211 Geneva, Switzerland}\\[0.1cm]
{\it $^b$ Skobeltsyn Institute of Nuclear Physics, Moscow State University, 119991 Moscow, Russia}\\

\vskip 3 truemm

\end{center}


\begin{abstract}
\noindent
The production of same-sign W-boson pairs from double parton scatterings (DPS) in proton-lead (\pPb) 
collisions at the CERN Large Hadron Collider is studied. The signal and background cross sections 
are estimated with next-to-leading-order perturbative QCD calculations using nuclear parton distribution 
functions for the Pb ion. At $\sqrtsnn$~=~8.8~TeV the cross section for the DPS process is about 150~pb, 
i.e. 600 times larger than that in proton-proton collisions at the same centre-of-mass energy and 1.5 times
higher than the \pPb~$\to$ W$^\pm$W$^\pm$+2-jets single-parton background. The measurement of such a process, 
where 10 events with fully leptonic W's decays are expected after cuts in 2~pb$^{-1}$, would constitute an
unambiguous DPS signal and would help determine the effective $\sigmaeff$ parameter characterising the
transverse distribution of partons in the proton.
\end{abstract}



\section{Introduction}
\label{sec:1}

The existence of multi-parton interactions (MPI) in high-energy hadronic collisions~\cite{MPI}
is a natural consequence of the fast increase of the parton flux at small parton fractional 
momenta $x$~$\equiv$~$p_{\rm parton}/p_{\rm hadron}$ and the requirement of unitarization of
the cross sections in perturbative quantum chromodynamics (pQCD)~\cite{Goebel:1979mi,Paver:1982yp,Humpert:1983pw,Mekhfi:1983az,Sjostrand:1987su}. 
Without MPI, the pQCD cross sections show a dramatic growth with increasing centre-of-mass 
energies leading to a violation of unitarity ($\sigma_{_{\rm parton-parton}}>\sigma_{\rm pp}$) 
at momentum transfers as large as $\cO($7~GeV$)$ in proton-proton (\pp) collisions at the 
Large Hadron Collider (LHC)~\cite{Grebenyuk:2012qp}. 
Basic experimental observations in proton-(anti)proton collisions 
-- such as the distributions of hadron multiplicities in minimum bias
collisions~\cite{GrosseOetringhaus:2009kz} and the underlying event 
activity in hard scattering interactions~\cite{Field:2010bc} -- 
can only be reproduced by Monte Carlo (MC) event generators such
as \pythia~\cite{Sjostrand:2004pf} and \herwig~\cite{herwig_mpi}, by including MPI contributions
which are modeled perturbatively using an impact-parameter description of the transverse parton 
profile of the colliding protons. 
Although MPI at semi-hard scales of a few~GeV are an unambiguous (experimental and theoretical)
source of particle production in high-energy hadronic collisions, 
the evidence for double parton scattering (DPS) processes 
producing two independently-identified hard particles 
in the same collision is currently based on relatively indirect measurements (see below).

The study of DPS processes provides not only valuable information on the spatial structure of the
hadrons~\cite{Diehl:2011yj} and on multi-parton correlations in the hadronic
wave-function~\cite{Calucci:2010wg}, but also a good experimental and theoretical control of the
DPS final-states is needed in searches of new physics at the LHC (such as e.g. 
Higgs~\cite{DelFabbro:1999tf,Hussein:2006xr,Bandurin:2010gn},
supersymmetry~\cite{Gaunt:2010pi}, and W$_{\rm L}$W$_{\rm L}$ scattering~\cite{WLWL}).
Various measurements exist in \pp\ and \ppbar\ collisions which are consistent with DPS contributions in
multi-jet final-states at $\sqrts$~=~63~GeV~\cite{Akesson:1986iv}, 630 GeV~\cite{Alitti:1991rd}, and
1.8~TeV~\cite{Abe:1993rv,Abe:1997xk}; in $\gamma$+3-jets events 
at $\sqrts$~=~1.8 TeV~\cite{Abe:1997bp} and 1.96~TeV~\cite{Abazov:2009gc}; and in (preliminary) W+2-jets
results at $\sqrts$~=~7~TeV~\cite{ATLAS-CONF-2011-160}. Such measurements show an excess of events in some
differential distributions with respect to the expectations from contributions from single-parton scatterings
(SPS) alone, although with uncertainties related to higher-order SPS corrections. The production of
same-sign WW production -- with a theoretical cross section with small uncertainties and a characteristic
final-state with like-sign leptons plus (large) missing transverse energy (\MET) from the undetected neutrinos -- 
has no SPS backgrounds at the same order in the strong coupling $\alphaS$, and has been proposed since various
years as a ``smoking gun'' process to univocally signal the occurrence of DPS in \pp\
collisions~\cite{Kulesza:1999zh,Cattaruzza:2005nu}.

In this paper we investigate DPS production of like-sign WW in proton-nucleus collisions
(\pA, where $A$ indicates the number of nucleons) at the CERN LHC. 
The larger transverse parton density in nuclei compared to protons results in enhanced
DPS cross sections coming from interactions where the two partons of the
nucleus belong to (i) the same nucleon and (ii) two different 
nucleons~\cite{Strikman:2001gz,DelFabbro:2003tj,Frankfurt:2004kn,Cattaruzza:2004qb,DelFabbro:2004md,Cattaruzza:2005nv,Treleani:2012zi,Blok:2012jr}:
\begin{equation} 
\sigmaDPS_{pA} = \sigmaDPSone_{pA} + \sigmaDPStwo_{pA}\,.
\label{eq:sigmaDPS_pA}
\end{equation} 
In Section~\ref{sec:2} we provide the generic form of both contributions in (\ref{eq:sigmaDPS_pA})
as a function of the corresponding single-parton proton-proton cross sections.
In Section~\ref{sec:3} we outline the theoretical setup used and in Section~\ref{sec:4} 
we quantify the cross sections for the production of same-sign W pairs and
associated backgrounds in proton-lead (\pPb) collisions in the range of LHC centre-of-mass (c.m.)
energies~\cite{Salgado:2011wc} and beyond~\cite{Assmann:1284326}, using next-to-leading-order (NLO)
calculations complemented with recent proton and nucleus parton distribution functions (PDF). 
Accounting for the leptonic decay branching ratios and applying standard kinematical requirements on the
final-state particles, we find about 10 DPS same-sign WW events expected in 2~pb$^{-1}$ integrated 
luminosity for \pPb\ at $\sqrtsnn$~=~8.8~TeV. The main conclusions of the work are summarised 
in Section~\ref{sec:5}. 

\section{Cross sections for double parton scattering in \pp\ and \pA\ collisions}
\label{sec:2}

\subsection{Generic hadron-hadron collisions}

In a generic hadron-hadron collision, the inclusive DPS cross section from two independent hard
parton subprocesses ($h h' \to a b$) has been derived in the momentum representation, taking into account the
transverse (impact parameter) distribution of the partons in the hadrons, using light-cone variables
and the same factorization approximations assumed for processes with a single hard
scattering~\cite{Paver:1982yp,Blok:2010ge,Diehl:2011tt,stir,Diehl:2011yj}. The DPS cross section 
can be written as a convolution of PDFs and elementary cross section summed over all involved partons
\begin{eqnarray} 
\sigmaDPS_{hh'\to a b} = & & \left(\frac{m}{2}\right) \sum \limits_{i,j,k,l} \int \Gamma_{h}^{ij}(x_1, x_2; {\bf b_1},{\bf b_2}; Q^2_1, Q^2_2)\nonumber\\
& &\times\,\hat{\sigma}^{ik}_{a}(x_1, x_1',Q^2_1) \, \hat{\sigma}^{jl}_{b}(x_2, x_2',Q^2_2)\\
& &\times\,\Gamma_{h'}^{kl}(x_1', x_2'; {\bf b_1} - {\bf b},{\bf b_2} - {\bf b}; Q^2_1, Q^2_2)\, dx_1 dx_2 dx_1' dx_2' d^2b_1 d^2b_2 d^2b\,.\nonumber
\label{eq:hardAB}
\end{eqnarray}
In this expression, $\Gamma_{h}^{ij}(x_1, x_2;{\bf b_1},{\bf b_2}; Q^2_1, Q^2_2)$ are double parton distribution
functions which depend on the longitudinal parton momentum fractions $x_1$ and $x_2$, and on the transverse position
${\bf b_1}$ and ${\bf b_2}$ of the two parton undergoing the hard processes 
at the scales $Q_1$ and $Q_2$, $\hat{\sigma}^{ik}_{a}$ and $\hat{\sigma}^{jl}_{b}$ are the
parton-level subprocess cross sections, and ${\bf b}$ is the impact parameter vector 
connecting the centres of the colliding 
hadrons in the trans\-verse plane. The factor $m/2$ appears due to the symmetry of the expression for
the interchange of $i$ and $j$ parton species: $m=1,2$ for indistinguishable and distinguishable 
parton subprocesses respectively. 
The double parton distribution functions $\Gamma_{h}^{ij}(x_1, x_2; {\bf b_1},{\bf b_2}; Q^2_1, Q^2_2)$ 
encode all the information of interest with regard to multiple parton interactions, and
it is typically assumed that they can be decomposed in terms of
longitudinal and transverse components
\begin{equation} 
\Gamma_{h}^{ij}(x_1, x_2;{\bf b_1},{\bf b_2}; Q^2_1, Q^2_2) 
 = D^{ij}_h(x_1, x_2; Q^2_1, Q^2_2) \, f({\bf b_1}) f({\bf b_2}),
\label{eq:DxF}
\end{equation} 
where $f({\bf b_1})$ describes the transverse parton density and is supposed to be an universal function for all
types of partons with the fixed normalization
\begin{eqnarray} 
\int f({\bf b_1}) f({\bf b_1 -b})d^2b_1 d^2b = \int t({\bf b})d^2b = 1,
\;\;\mbox{ with the overlap function}\;\; t({\bf b}) = \int f({\bf b_1}) f({\bf b_1 -b})d^2b_1\,.
\label{eq:f}
\end{eqnarray} 
Making the further assumption that the longitudinal components $D^{ij}_h(x_1, x_2; Q^2_1, Q^2_2)$
reduce to the ``diagonal'' product of two independent single-parton distribution functions, 
\begin{eqnarray} 
D^{ij}_h(x_1, x_2; Q^2_1, Q^2_2) = D^i_h (x_1; Q^2_1) \,D^j_h (x_2; Q^2_2),
\label{eq:DxD}
\end{eqnarray}
the cross section of double parton scattering can be finally expressed in the simple generic form
\begin{equation} 
\sigmaDPS_{hh'\to a b} = \left(\frac{m}{2}\right) \frac{\sigmaSPS_{hh'\to a} \cdot \sigmaSPS_{hh'\to b}}{\sigmaeff}\,,
\label{eq:doubleAB}
\end{equation}
where $\sigmaSPS$ is the standard inclusive single-hard scattering cross section, computable perturbatively to
a given order in $\alphaS$,
\begin{eqnarray} 
\sigmaSPS_{hh' \to a} & = & \sum \limits_{i,k} \int D^{i}_h(x_1; Q^2_1) f({\bf b_1})\,
\hat{\sigma}^{ik}_{a}(x_1, x_1') \times\, D^{k}_{h'}(x_1'; Q^2_1)f( {\bf b_1} - {\bf b}) dx_1 dx_1' d^2b_1
d^2b\nonumber \\
& = & \sum \limits_{i,k} \int D^{i}_h(x_1; Q^2_1) \,\hat{\sigma}^{ik}_{a}(x_1, x_1') \,D^{k}_{h'}(x_1'; Q^2_1) dx_1 dx_1'\,,
\label{eq:hardS}
\end{eqnarray}
and $\sigmaeff$ is a normalization cross section associated with the effective transverse 
overlap area (\ref{eq:f}) of the partonic correlations that produce the DPS process:
\begin{eqnarray} 
\sigmaeff = \left[ \int d^2b \, t^2({\bf b})\right]^{-1}\,.
\label{eq:sigmaeff1}
\end{eqnarray} 
In \pp\ collisions, the approximate range of numerical values of this effective cross section has been
obtained empirically from fits to data~\cite{Akesson:1986iv,Alitti:1991rd,Abe:1993rv,Abe:1997xk,Abe:1997bp,Abazov:2009gc,ATLAS-CONF-2011-160}:
\begin{eqnarray} 
\sigmaeffpp\approx 13\pm2\;\mbox{~mb.}
\label{eq:sigmaeff}
\end{eqnarray} 

The validity of the simplifying assumptions leading to Eq.~(\ref{eq:doubleAB}) in combination 
with the interpretation of $\sigmaeff$ as a measure of the effective transverse parton interaction 
area given by Eq.~(\ref{eq:sigmaeff1}) 
-- which are quite customary and economic expressions from a computational point of view -- is 
the object of current
revision~\cite{Blok:2010ge,Diehl:2011tt,stir,Ryskin:2011kk,Diehl:2011yj,Blok:2011bu,Gaunt:2012wv,Manohar:2012pe,Ryskin:2012qx,Gaunt:2012dd}.
In particular, the presence of a correlation term in the two-parton distributions, now neglected in Eq.~(\ref{eq:DxD}),
results in a decrease of the effective cross section $\sigmaeff$ (i.e. in an {\it increase} of the final DPS cross section) 
with the growth of the hard scale~\cite{Snigirev:2010tk,Ryskin:2011kk,flesburg}, while the dependence of
$\sigmaeff$ on the total energy at fixed scales is rather weak~\cite{flesburg}. Thus, 
Eq.~(\ref{eq:doubleAB}), which will be employed hereafter in this work, should actually provide a conservative
estimate of the DPS cross section for the process under consideration.

\subsection{Proton-nucleus collisions}

In proton-nucleus collisions, the parton flux is enhanced by the number $A$ of nucleons in the nucleus 
and -- modulo small (anti)shadowing effects in the nuclear PDF~\cite{Armesto:2006ph}, see below -- the single-parton cross section 
is simply expected to be that of proton-proton collisions (or, more
exactly, that of proton-nucleon collisions \pN\ with N=p,n including protons and neutrons with their 
appropriate relative fraction) scaled by the factor $A$~\cite{d'Enterria:2003qs}, i.e.
\begin{eqnarray} 
\sigmaSPS_{pA \to a b} = \sigmaSPS_{pN \to a b} \int \TpA({\bf r}) \, \dtwor = A \cdot \sigmaSPS_{pN \to a b}\,.
\label{eq:sigmaSPSpA}
\end{eqnarray} 
Here, $\TpA({\bf r})$ is the standard nuclear thickness function, analogous to $f({\bf b})$ in Eqs.~(\ref{eq:DxF}) and
(\ref{eq:f}), as a function of the impact parameter ${\bf r}$ between the colliding proton and nucleus, 
given by an integral of the density function over the longitudinal direction $\TpA({\bf r}) = \int \rho_A({\bf r})\,dz$, 
normalised to $\int \TpA({\bf r})\,\dtwor = A$.

The corresponding DPS \pA\ cross section is the sum of the two terms of Eq.~(\ref{eq:sigmaDPS_pA}):
\begin{itemize}
\item The first term is just the DPS cross section in \pN\ collisions 
similarly multiplied by $A$: 
\begin{eqnarray} 
\sigmaDPSone_{pA \to a b} = A \cdot \sigmaDPS_{pN \to a b}\,,
\label{eq:doubleAB1}
\end{eqnarray} 
\item the second contribution, 
for which interactions with partons from two different nucleons are involved in the scattering, 
involves the square of $\TpAsq$~\cite{Strikman:2001gz}
\begin{eqnarray} 
\label{eq:doubleAB2}
\sigmaDPStwo_{pA \to a b} = \sigmaDPS_{pN \to a b} \cdot \sigmaeffpp \cdot \rm{F}_{pA},\\
\mbox{ with } \; \rm{F}_{pA} = \frac{A-1}{A} \int \TpAsq({\bf r})\,\dtwor,
\label{eq:TpAsq}
\end{eqnarray} 
which assumes again that the double PDF of the nucleons are factorised in both longitudinal and
transverse components as in Eqs.~(\ref{eq:DxF}) and (\ref{eq:DxD}). The factor $(A-1)/A$ is introduced to take
into account the difference between the number of nucleon pairs and the number of different nucleon pairs.
The effect of extra ``non-diagonal'' interference terms (which cannot be expressed in terms of parton densities alone)
computed in the case of light nuclei in~\cite{Treleani:2012zi} would lead, together with other correlations
discussed above, to an increase of the value of $\sigmaDPStwo_{pA \to a b}$, but we believe that these are
effectively covered by the current uncertainty $\sigmaeffpp$, Eq.~(\ref{eq:sigmaeff}).
\end{itemize}

Thus, adding (\ref{eq:doubleAB1}) and (\ref{eq:doubleAB2}) the inclusive cross section of a DPS process with
two hard parton subprocesses $a$ and $b$ in a \pA\ collision can be written as
\begin{eqnarray} 
\sigmaDPS_{pA\to a b} = A \,\sigmaDPS_{pN \to a b}\left[1+\frac{1}{A}\, \sigmaeffpp \,\rm{F}_{pA}\right]\,,
\label{eq:doublepA}
\end{eqnarray}
which is enhanced by the factor in parentheses compared to the corresponding DPS cross section in
proton-proton (or, more exactly, \pN) collisions. 
Let us evaluate this factor in the case of proton-lead collisions.
In the simplest approximation that the nucleus has a spherical form with uniform
nucleon density of radius $R_A=r_0 A^{1/3}$ with $r_0=1.25$~fm, the integral 
of the nuclear thickness factor (\ref{eq:TpAsq}) is $\rm{F}_{pA} = 9 A(A-1)/(8 \pi R_A^2)$~=~31.5~mb$^{-1}$
for a Pb nucleus ($A$~=~208).
If, instead of the hard-sphere approximation, one directly evaluates the integral 
using the standard Fermi-Dirac 
spatial density for the lead nucleus ($R_A$~=~6.36~fm and surface thickness $a$~=~0.54~fm)~\cite{deJager} 
one obtains $\rm{F}_{pA}$~=~30.4~mb$^{-1}$. Thus, for the value of $\sigmaeffpp$ determined
experimentally, Eq.~(\ref{eq:sigmaeff}), the DPS enhancement factor (\ref{eq:doublepA}) in \pPb\ compared to
A-scaled \pN\ collisions is of order
$[ 1 + \sigmaeffpp \rm{F}_{pA}/A ] \simeq 3$. 

The final formula for DPS in proton-nucleus collisions can be written as a function of the elementary
proton-nucleon single-parton cross sections, combining Eqs.~(\ref{eq:doubleAB}) and (\ref{eq:doublepA}), as
\begin{eqnarray} 
\sigmaDPS_{pA\to a b} = \left(\frac{m}{2}\right) \frac{\sigmaSPS_{pN \to a} \cdot \sigmaSPS_{pN \to b}}{\sigmaeffpA}\,,
\label{eq:sigmapPbDPSWW}
\end{eqnarray}
with the normalization effective cross section amounting to
\begin{eqnarray} 
\sigmaeffpA = \frac{\sigmaeffpp}{A+\sigmaeffpp\,\rm{F}_{pA}} = 21.5 \pm 1.1 \mbox{ $\mu$b}
\label{eq:sigmaeffpPb}
\end{eqnarray}
where the last equality holds for \pPb\ using $A$~=~208, $\sigmaeffpp =$~13~$\pm$~2~mb and $\rm{F}_{pA}$~=~30.4~mb$^{-1}$. 
In summary, the DPS cross sections are enhanced by a factor of $\sigmaeffpp/\sigmaeffpA \approx 3\,A \approx$~600 in
proton-lead compared to proton-proton (or, in general, nucleon-nucleon) collisions, 
i.e. they are a factor of 3 higher than the naive expectation based on the $A$-scaling of the
single-parton cross sections, Eq.~(\ref{eq:sigmaSPSpA}). Thus, in general the significance 
for any DPS measurement in \pA\ collisions, 
$\Sig$~=~N$^{^{\rm {\tiny DPS}}}_{_{\rm {\tiny pA}}}/\sqrt{\rm N^{^{\rm {\tiny SPS}}}_{_{\rm {\tiny pA}}}}$,
will be enhanced by a factor of $3 \sqrt{A}$ compared to \pp\ collisions, i.e. by a factor of
$\sim$40 for \pPb. One can thus exploit such a large expected DPS signal over the SPS background in proton-nucleus
collisions to study double-parton scattering in detail and in particular to
determine the value of $\sigmaeffpp$, independently of other DPS measurements in \pp\ collisions --
given that the parameter F$_{pA}$ in Eq.~(\ref{eq:sigmaeffpPb}) depends on the comparatively better known
transverse density profile of nuclei.

Although in this paper we will be computing the proton-nucleon single-parton cross sections using NLO
calculations that take directly into account the proper combination of \pp\ and \pn\ collisions 
-- i.e. we will be using directly Eq.~(\ref{eq:sigmapPbDPSWW}) with $\sigmaSPS_{pN \to a}=\sigmaSPS_{pA \to a}/A$ 
calculated using nuclear PDFs for the Pb ion to obtain the corresponding DPS cross sections -- we close this Section by writing
explicitly the generic \pA\ DPS expression including the individual \pp\ and \pn\ 
collisions for a generic nucleus of proton number Z and neutron number N~=~A~--~Z: 
\begin{eqnarray} 
\sigmaDPS_{pA\to a b} & = &\left(\frac{m}{2}\right) \;\frac{Z \cdot\sigmaSPS_{pp\to a} \cdot \sigmaSPS_{pp\to b}
+ N\cdot\sigmaSPS_{pn\to a} \cdot \sigmaSPS_{pn\to b}}{\sigmaeffpp} \nonumber \\
& + & \left(\frac{m}{2}\right)\mathrm{F}_{pA} \; \Big[ Z(Z-1)/(A(A-1))\cdot\sigmaSPS_{pp\to a} \cdot \sigmaSPS_{pp\to b}
+ Z\cdot N/(A(A-1))\cdot\sigmaSPS_{pp\to a} \cdot \sigmaSPS_{pn\to b} \\
& + & Z\cdot N/(A(A-1))\cdot\sigmaSPS_{pn\to a} \cdot \sigmaSPS_{pp\to b} + N(N-1)/(A(A-1))\cdot\sigmaSPS_{pn\to a} \cdot \sigmaSPS_{pn\to b}\Big]\nonumber 
\end{eqnarray}


\section{Theoretical setup} 
\label{sec:3}

The interest of like-sign W-boson pair production as a signature of DPS in high-energy hadronic collisions is
three-fold. 
First, all the relevant production cross sections can be computed perturbatively at NLO accuracy
and have small theoretical uncertainties.
Second, the W$^\pm$ production cross sections at the LHC are the highest for any electroweak particle
(Z and Drell-Yan production are comparatively smaller), resulting in a potentially 
observable DPS WW cross section. 
Third, in single-parton scatterings, the same-sign SPS processes are suppressed
since the lowest order at which two same-sign W bosons can be produced is with two jets (WWjj)~\cite{Kulesza:1999zh},
$q\,q \to $W$^+$W$^+ \,q'\,q'$ with $q=u,c,\ldots$ and $q'=d,s,\ldots$
which have leading contributions of $\cO(\alpha_{\rm s}^2\alpha_{\rm w}^2)$ for the mixed strong-electroweak
diagrams, and of $\cO(\alpha_{\rm w}^4)$ for the pure electroweak (vector-boson fusion) processes,
where $\alpha_{\rm w}$ is the electroweak coupling.

We compute the same-sign DPS signal, $\sigmaDPS_{pPb\to WW}$, via Eqs.~(\ref{eq:sigmapPbDPSWW})-(\ref{eq:sigmaeffpPb}) with $m$~=~1.
The single-parton W cross section $\sigmaSPS_{pN \to W}$ entering in the DPS expression, 
as well as the SPS same-sign WWjj background,
are computed at NLO accuracy with the \mcfm\ code~\cite{mcfm,Campbell:2011bn} (version 6.2) using 
the central NLO sets of the CT10 PDF~\cite{Lai:2010vv} for the proton and using the EPS09 nuclear
PDF~\cite{Eskola:2009uj} for the Pb ion. 
The NLO calculations\footnote{Next-to-NLO calculations, available for the single-W production, increase by a
few percent the pQCD cross sections and further improve the data-theory agreement.} used here reproduce well
the experimental single-W~\cite{Aad:2010yt,Khachatryan:2010xn,Chatrchyan:2012nt} and
W$^+$W$^-$~\cite{Chatrchyan:2011tz,Aad:2011kk,ATLAS:2012mec,WWcms2012} 
production cross sections measured in \pp\ collisions at $\sqrtsnn$~=~7~TeV, as well as the single-W
cross sections in \pp\ and \PbPb\ at $\sqrtsnn$~=~2.76~TeV~\cite{Chatrchyan:2012nt}.

The use of nuclear PDF (EPS09) for the lead ion not only takes into account nuclear ``shadowing''
modifications of the bound relative to free nucleons, but it also properly accounts for the different
isospin (u- and d-quark) content of the lead ion given by its different proton (Z = 81) and neutron (N = 127) numbers. 
Such an effect is important in the case of isospin-sensitive particles like the W boson, and explains 
the relatively enhanced  W$^-$  (depleted W$^+$) cross sections measured in \PbPb\ with respect to \pp\ 
collisions at $\sqrtsnn$~=~2.76~TeV~\cite{Chatrchyan:2012nt}.
At 8.8~TeV, the use of the EPS09 nuclear PDF results in a modification of the total W$^+$ (W$^-$) production
cross section by about -7\% (+15\%) compared to those obtained for \pp\ collisions using the proton CT10
PDF. The largest impact in the final cross sections is due to the proton-versus-lead isospin differences
since nuclear (anti)shadowing effects alone decrease the integral W$^+$ and W$^-$ yields respectively 
by 4.5\% and 3\% only~\cite{Paukkunen:2010qg}.

All numerical results are evaluated using the latest standard model input parameters for particle masses,
widths and couplings~\cite{PDG}. 
The single-W$^\pm$ production is computed fixing the renormalization and factorization scales at $\mu = \mu_F
= \mu_R$~=~$m_W$, whereas we use $\mu$~=~2$m_W$ for the W$^+$W$^-$ cross section (shown here only as a
reference). 
The background WWjj cross sections in \mcfm\ are formally computed at LO, but
detailed studies~\cite{Melia:2010bm,Melia:2011gk,Jager:2011ms} have shown that the extra NLO corrections 
can be accounted for by setting the renormalization and factorization scales to $\mu$~=~150~GeV. 
Indeed, for such a scale choice the W$^+$W$^+$+2-jet inclusive LO and NLO cross sections are found to 
nearly coincide in \pp\ at $\sqrts$~=~14 TeV ($\sigma_{\rm LO}$~=~2.4~fb and $\sigma_{\rm NLO}$~=~2.5~fb)~\cite{Melia:2010bm}. 
Same-sign W-pairs can also be produced via vector-boson-fusion (VBF) processes~\cite{Jager:2009xx,Denner:2012dz}, 
which are not included in the \mcfm\ package. We have determined the VBF WWjj contribution at NLO 
using \vbfnlo~\cite{vbfnlo,Arnold:2012xn} (version 2.6) with the same set of parameters 
used for \mcfm, except that the theoretical scales are set to the momentum transfer of the exchanged W,\,Z boson
($\mu^2 = t_{W,Z}$). Since \vbfnlo\ is not interfaced to nuclear PDFs, we use CT10 for both the proton and the
Pb ion. This choice introduces a small difference (below 5\%) in the VBF cross sections according to the results
obtained running a similar weak-boson fusion process in \mcfm\ (VBF Higgs$\to$W$^+$W$^-$ production) with CT10 and EPS09.
The pure \vbfnlo\ electroweak contributions to the total WWjj cross sections are as large (or even
slightly larger than) the QCD (quark- and gluon-mediated) ones obtained with \mcfm\ (Table~\ref{tab:1}).

The uncertainties of the SPS NLO single-W cross sections amount to about $\pm$10\% as obtained by taking into
account the EPS09 nPDF eigenvector set (the proton PDF uncertainties are much lower in the relevant regions 
of parton fractional momentum and virtuality), and by
independently varying 
the $\mu_F$ and $\mu_R$ scales within a factor of two. The quoted uncertainties for the QCD WWjj cross
sections are those from the full-NLO calculations~\cite{Melia:2010bm}. The uncertainties for the VBF cross
sections are much smaller as they do not involve any gluons in the initial state. In the case of the DPS cross
section, the SPS uncertainties are added in quadrature with the $\pm$15\% uncertainty of $\sigmaeffpp$,
Eq.~(\ref{eq:sigmaeff}), and result in a total uncertainty of about $\pm$18\% in the final cross section.


\section{Results}
\label{sec:4}

Table~\ref{tab:1} collects the cross sections and associated uncertainties for the relevant individually computed processes 
for \pPb\ collisions at two c.m. energies $\sqrtsnn$~=~5.0~TeV (corresponding to the first \pPb\ run in 2013)
and 8.8~TeV (nominal energy). We note that, in general, the SPS cross sections presented here are larger than those
obtained for \pp\ collisions (scaled by $A$) in previous calculations~\cite{Gaunt:2010pi}, since the latter
have been computed only at LO accuracy\footnote{For single-W, the K-factor 
$K = \sigma_{_{\rm NLO}}/\sigma_{_{\rm LO}}\approx$~1.15 results in a 30\% increase of the DPS WW
cross section compared to the LO DPS estimate.}.

\begin{table}[htbp]
\caption{Inclusive cross sections for single-W and double-W scatterings in \pPb\ collisions at two
 c.m. energies. The single-parton-scattering cross sections are obtained at NLO 
with \mcfm\ and \vbfnlo\ for the processes quoted. The last column lists the same-sign
double-parton-scattering (DPS) cross sections (sum of positive and negative W pairs) 
obtained with Eqs.~(\ref{eq:sigmapPbDPSWW})-(\ref{eq:sigmaeffpPb}).
\label{tab:1}}
\vspace{-0.5cm}
\begin{center}
\begin{tabular}{l|cc|c|cc|c}\hline
\hspace{-0.5mm} \pPb\ final-state: \hspace{-0.5mm} & \hspace{-0.5mm} W$^+$ \hspace{-0.8mm} & 
\hspace{-0.8mm} W$^-$ \hspace{-0.5mm} & \hspace{-0.5mm} W$^+$W$^-$ \hspace{-0.5mm} & \hspace{-0.5mm} W$^+$W$^+$jj (QCD) \hspace{-1.0mm} & 
\hspace{-1.0mm} W$^+$W$^+$jj (VBF) \hspace{-0.5mm} & \hspace{-0.5mm} W$^\pm$W$^\pm$ (DPS) \hspace{-0.5mm} \\
\hspace{-0.5mm} Code (\ttt{process \#}): \hspace{-0.5mm} & \hspace{-0.5mm} \mcfm\ (\ttt{1}) \hspace{-0.8mm} & 
\hspace{-0.8mm} \mcfm\ (\ttt{6})\hspace{-0.5mm} & \hspace{-0.5mm} \mcfm\ (\ttt{61}) \hspace{-0.5mm} & 
\hspace{-0.5mm} \mcfm\ (\ttt{251}) \hspace{-0.5mm} & \hspace{-0.5mm} \vbfnlo\ (\ttt{250}) \hspace{-0.5mm} & 
\hspace{-0.5mm} \small{Eq.~(\ref{eq:sigmapPbDPSWW})} \hspace{-0.5mm}\\
\hspace{-0.5mm} Order $(\sigma$ units): \hspace{-0.5mm} & \hspace{-0.5mm} \small{NLO ($\mu$b)} \hspace{-0.8mm} & 
\hspace{-0.8mm} \small{NLO ($\mu$b)} \hspace{-0.5mm} & \hspace{-0.5mm} \small{NLO (nb)} \hspace{-0.5mm} & 
\hspace{-0.5mm} \small{'NLO' (pb)} \hspace{-0.5mm} & \hspace{-0.5mm} \small{NLO (pb)} \hspace{-0.5mm} & 
\hspace{-0.5mm} \small{(pb)} \hspace{-0.5mm}\\\hline
\hspace{-0.5mm} $\sqrtsnn$~=~5.0~TeV \hspace{-0.5mm} & 
\hspace{-0.5mm} 6.85 $\pm$ 0.68 \hspace{-0.8mm} & \hspace{-0.8mm} 5.88 $\pm$ 0.59 \hspace{-0.5mm} & 
\hspace{-0.5mm} 5.48 $\pm$ 0.56 \hspace{-0.5mm} & \hspace{-0.5mm} 12.1 $\pm$ 1.2 \hspace{-0.5mm} & 
\hspace{-0.5mm} 12.4 $\pm$ 0.6 \hspace{-0.5mm} & \hspace{-0.5mm} 44. $\pm$ 8. \hspace{-0.5mm}\\

\hspace{-0.5mm} $\sqrtsnn$~=~8.8~TeV \hspace{-0.5mm} & 
\hspace{-0.5mm} 12.6 $\pm$ 1.3 \hspace{-0.8mm} & \hspace{-0.8mm} 11.1 $\pm$ 1.1 \hspace{-0.5mm} &
\hspace{-0.5mm} 13.0 $\pm$ 1.3 \hspace{-0.5mm} & \hspace{-0.5mm} 40.4 $\pm$ 4.0 \hspace{-0.5mm} & 
\hspace{-0.5mm} 51.8 $\pm$ 2.0 \hspace{-0.5mm} & \hspace{-0.5mm} 152. $\pm$ 27. \hspace{-0.5mm}\\\hline
\end{tabular}
\end{center}
\end{table}

\begin{figure}[htpb]
\centering
\epsfig{figure=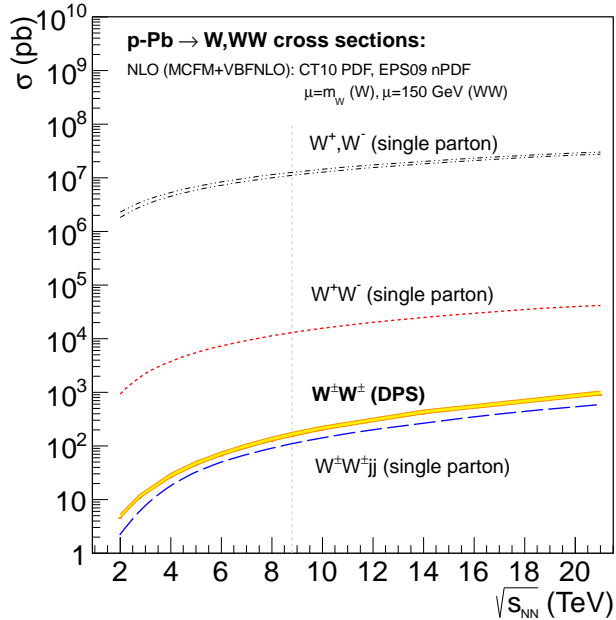,width=0.5\columnwidth}
\caption{Total production cross sections for single-W 
and W-pair boson(s) in \pPb\ collisions as a function of c.m. energy, for single-parton and double-parton
(DPS) scatterings. The W$^\pm$W$^\pm$ labels indicate that W$^+$W$^+$ and W$^-$W$^-$ cross sections have been added.
The width of the DPS curve indicates its associated uncertainty (dominated by $\sigmaeff$).
The dashed vertical line indicates the nominal $\sqrtsnn$~=~8.8~TeV nucleon-nucleon c.m. energy 
in p-Pb collisions at the LHC.
\label{fig:1}}
\end{figure}

Figure~\ref{fig:1} shows the resulting total cross sections for all relevant processes in \pPb\ in the range
of c.m. energies in the nucleon-nucleon system of $\sqrtsnn \approx$~~2--20~TeV. 
The nominal \pPb\ LHC energy is 8.8~TeV (dashed vertical line in the plot) but we have extended the
calculations up to 2.5 times this value into the range reachable in a proposed future high-energy upgrade of
the collider~\cite{Assmann:1284326}. The upper curve in the plot corresponds to the SPS cross sections
for W$^+$ and W$^-$ production (at 8.8~TeV the former is about 12\% higher than the latter). 
The second curve corresponds to the unlike-sign W$^+$W$^-$ SPS cross sections, which is added to the plot
for comparison purposes. The third curve corresponds to the same-sign WW signal DPS cross section
which rises from $\sigmaDPS_{pPb\to WW}\approx$~5~pb up to about 1~nb in the considered range of
c.m. energies. The last curve corresponds to the SPS background $\sigmaSPS_{pPb\to WWjj}$, 
obtained adding the QCD (\mcfm) and weak-boson-fusion (\vbfnlo) cross sections for negative and positive W
pairs, which is found to be about a factor of 1.5 lower than the DPS signal.

The cross sections plotted in Figure~\ref{fig:1} are total inclusive ones and do not include W decays, nor
cuts on any of the final-state particles. The W leptonic fraction ratios, W$\to \ell^{\pm} \nu_{\ell}$ ($\ell$ = e, $\mu$,
$\tau$), amount to 1/9 for each lepton flavour, the other 2/3 being due to hadronic quark-antiquark decays.
Taking into account the standard electron and muon decay modes measured for both W-bosons ($ee,\mu\mu,e\mu,\mu e$)
would therefore reduce the WW yields by a factor of $\bR_{_{\rm \ell\ell}} = (4/9^2)\approx$~1/20, although
including $e^\pm$ and $\mu^\pm$ from leptonic tau-decays in the WW$\to e\tau,\mu\tau,\tau\tau$ 
final-states would decrease the corresponding branching ratio only to $\bR_{_{\rm \ell\ell}} \approx$~1/16.
A more aggressive scenario combining the (W$\to \mu$,$e$) and (W$\to$jj) decays, as done at
Tevatron~\cite{Abazov:2008yg,Aaltonen:2009vh,Aaltonen:2010rq} and also at the LHC~\cite{WWcms2012}, would result in a
reduction of the visible cross section by a branching factor of $\bR_{_{\rm \ell+2j}}=(2/9\cdot
4/3)\approx$~1/3.3, though this would require the determination of the sign of the charge of the jets which 
is not obvious a priori (although see e.g.~\cite{Krohn:2012fg} for recent developments).
In order to conservatively estimate the expected yields measured at the LHC, we consider 
the purely leptonic decay branching ratios and we include in the generator-level \mcfm\ calculations 
typical ATLAS and CMS fiducial requirements for the decay leptons: $|\eta^{\ell}|<$~2.4,
$p_T^{\ell}>$~20~GeV/c and $\MET>$~20~GeV~\cite{Aad:2010yt,Khachatryan:2010xn}. Such 
a kinematical selection reduces the visible WW cross section by about a factor of two which, combined
with small reconstruction inefficiencies of the final-state particles, 
would result in a factor of $\varepsilon \approx$~1/30 reduction of the WW yields at the LHC. 
We do not consider here other possible backgrounds ($\ttbar$, W,Z+jets, WZ,ZZ production ...) which are commonly
removed applying a jet-veto requirement and/or extra criteria on the invariant masses of the two leptons.

In conclusion, for a DPS cross section of $\sigmaDPS_{pPb\to WW}\approx$~150 pb at $\sqrtsnn$~=~8.8~TeV the
total number of events expected in one \pPb\ run is $\cN = \sigmaDPS_{pPb\to WW}\cdot\LumiInt\cdot\varepsilon \approx$~1--10
for an integrated-luminosity of $\LumiInt$~=~0.2--2~pb$^{-1}$, where the first value is the nominal (but
conservative) luminosity~\cite{Salgado:2011wc} and the second one assumes that one can use the same proton
beam intensity and/or emittance as in the current \pp\ running. 
We note also that the \pPb\ instantaneous luminosities at the LHC, at variance with the \pp\ operation mode, 
result in a very small event pileup~\cite{d'Enterria:2009er} and make the measurement accessible 
without complications from overlapping \pPb\ collisions occurring simultaneously in the same bunch crossing.


\section{Conclusions}
\label{sec:5}

The formalism of double parton scattering (DPS) in high-energy proton-nucleus collisions 
has been reviewed and a simple generic formula has been derived for the computation of the 
corresponding DPS cross sections as a function of (i) the single-parton cross sections in 
proton-proton collisions, and (ii) the effective $\sigmaeff$ parameter describing the transverse density of
partons in the proton. The DPS cross sections in \pPb\ are found to be enhanced by a factor of $3\,A \approx$~600
compared to those in \pp\ collisions at the same energy. 
The significance of the DPS measurements over the expected backgrounds 
in \pPb\ is a factor of $3\,\sqrt{A} \approx$~40 higher than that in \pp\ collisions. 
As a particular case, we have studied same-sign W-boson pair production in proton-lead 
collisions at LHC centre-of-mass energies, using NLO predictions -- \mcfm\ with nuclear PDFs for the QCD 
processes, and \vbfnlo\ for the electroweak backgrounds~-- for the production of single inclusive W and 
for the same-sign WWjj backgrounds. At the nominal $\sqrtsnn$~=~8.8~TeV energy, the DPS cross section for
like-sign WW production is about 150~pb, i.e. 600 times larger than that in proton-proton collisions at the
same centre-of-mass energy and 1.5 times higher than the single-parton same-sign WW+2-jets background. The
measurement of such a process, where 10 events with fully leptonic W's decays are expected after cuts in
2~pb$^{-1}$, would constitute an unambiguous DPS signal at the LHC, and would help determine the $\sigmaeff$
parameter characterising the effective transverse parton area of hard interactions in hadronic collisions.

\section*{Acknowledgments}
We are grateful to Carlos Salgado for discussions and valuable feedback on the \mcfm\ interface with the EPS09
nuclear PDF, as well as to Igor~Lokhtin for useful discussions. This work is partly supported by Russian
Foundation for Basic Research (RFBR) grant No. 10-02-93118 and the CERN-RFBR Joint Research Grant No. 12-02-91505.



\end{document}